\documentclass[]{raa}            
\usepackage{graphicx,times}
\usepackage{natbib}

\providecommand{\bm}[1]{\mbox{\boldmath$#1$\unboldmath}}

\begin{document}

\title{The supernova remnant CTB 37B and its associated magnetar CXOU J171405.7-381031: evidence for a
magnetar-driven remnant}

 \volnopage{ {\bf } Vol.\ {\bf } No. {\bf }, 000--000}
   \setcounter{page}{1}

   \author{J. E. Horvath
      \inst{1}
   \and M. P. Allen
      \inst{2}
   }

    \institute{Instituto de Astronomia, Geof\'\i sica e Ci\^encias
Atmosf\'ericas - Universidade de S\~ao Paulo, Rua do Mat\~ao,
1226, 05508-900, Cidade Universit\'aria, S\~ao Paulo SP, Brazil;
{\it foton@astro.iag.usp.br}
        \and
             CEFET S\~ao Paulo,
Rua Pedro Vicente 625, 01109-010, Canind\'e, S\~ao Paulo SP,
Brazil}

\vs \no
   {\small Received [year] [month] [day]; accepted [year] [month] [day]
   }

\abstract{We discuss in this Letter the association of the
candidate magnetar CXOU J171405.7-381031 with the supernova
remnant CTB 37B. The recent detection of the period derivative of
the object allowed an estimation of a young characteristic age of
only $\sim 1000 yr$. This value is too small to be compatible even
with the minimum radius of the remnant $\geq 10 pc$, the value
corresponding to the {\it lower} limit of the estimated distance
of $10.2 \pm 3.5 kpc$, unless the true distance happens to be even
smaller than the lower limit. We argue that a consistent scenario
for the remnant origin, in which the latter is powered by the
energy injected by a young magnetar, is indeed more accurate to
explain the young age, and points out to its non-standard (i.e.
magnetar-driven) nature. }

 $\cdots\cdots$ 

   \authorrunning{J.E. Horvath \& M.P. Allen }            
   \titlerunning{CTB 37B is a magnetar-driven remnant}  
   \maketitle


%
%
\section{Introduction}           
\label{sect:intro}

Firmly placed inside the population of young neutron stars, the
sub-class of high-B field sources, currently composed of the
Anomalous X-ray Pulsars and the Soft-Gamma Repeaters has been well
identified and intensively studied over the last decades (Woods \&
Thompson 2006). However, there are many puzzles concerning their
progenitors and birth events. Given their short characteristic
ages ($\tau = {P\over{2 {\dot P}}}$), they should still be
associated with SN remnants. Based on positional coincidence,
similarity of ages and other observable features only a few
associations remain undisputed (Gaensler et al. 2001, Ankay et al.
2001, Mardsen et al. 2001, Allen \& Horvath 2004). This topic may
hide clues for a proper understanding of their nature and
evolution, towards an explanation of the apparent high value of
the magnetar magnetic fields.

While the evolution of conventional SNR has been widely discussed
in the literature (see, for instance, Truelove \& McKee 1999), it
has been argued that the very formation of a highly magnetized
object on short timescales should lead to a dynamical behavior of
the remnant, hereby termed {\it magnetar-driven SNR}. This
modified dynamics , in turn, has implications for the proposed
associations and should be taken into account for a consistent
picture. We shall review the fundamental ideas in section 2 and,
for the sake of definiteness, and apply them to the recently
proposed candidate CXOU J171405.7-381031 in CTB 37B, identified as
a magnetar by two different groups (Sato et al. 2010, Halpern \&
Gotthelf 2010b). We will argue that a short derived characteristic
age $\sim 1000 yr$ for the magnetar poses problems for a
``standard'' (i.e. not driven by energy injection of the
magnetar), and actually supports the magnetar-driven picture.

\section{Energy injection in magnetar-driven SNR}
\label{sect:Energy}

In spite of several decades of work, the problem of gravitational
collapse explosions is not solved (Burrows \& Nordhaus 2009). A
sequence of events along the collapse has been well established,
but substantial difficulties remain concerning the detailed
mechanism(s) of the explosion, and the role (if any) of the formed
compact object. Several studies of the expansion of remnants in
different ISM were performed over the years, just assuming that
the explosion is successful, and an energy of $\sim \, 10^{51} \,
erg$ (Hamuy 2003) is released in a point-like region. Of course if
the central object is to become decoupled for the outgoing gas,
this approach makes sense. However, such vision was first
challenged by Ostriker and Gunn (1971) when they postulated that
the rotation energy of a central pulsar may drive the supernovae.
Later it became clear that this phenomenon is too slow to power
explosions, but many features of it still remain in the so-called
magnetodynamical mechanisms (Moiseenko, Bisnovatyi-Kogan \&
Ardeljan 2010) . A recent work addressing the lightcurve and
energetics issue for the specific case of magnetars can be found
in Woosley (2010).

The discovery of superstrong magnetic fields posed yet another
related problem to the explosion scenario. If the field had to be
amplified from an initial seed, then the pre-supernova progenitors
had to possess a suitable distribution of both magnetic field and
angular momentum. If the initial rotation is not fast enough, the
amplification is quenched (Thompson, Chang, \& Quataert 2004) .
This leads to an idea that fast rotating magnetars should be born
to grow their fields, although their braking is efficient in the
aftermath following their birth. Thus, if a {\it magnetar} is in
turn formed inside the remnant immediately after the explosion by
dynamo amplification, the injection of energy (much in the same
way as Ostriker and Gunn 1971 envisioned) is inevitable, and an
initial energy loss $L_{0} = 3.85 \times 10^{47}
{\biggl({\frac{B}{10^{14} \, G}}\biggr)}^{2} {\biggl({\frac{1 \,
ms}{P_{0}}}\biggr)}^{4} \, erg s^{-1}$ and initial timescale for
deceleration $\tau_{0} \, = \, 0.6 {\biggl({\frac{10^{14} \,
G}{B}}\biggr)}^{2} {\biggl({\frac{P_{0}}{1 \, ms}}\biggr)}^{2} \,
d$ can be defined for this process (Allen \& Horvath 2004). These
estimates are strictly valid for constant values of the magnetic
field, whereas it is clear that the very phenomenon of the field
growth is operating here. However, in the absence of a full
detailed treatment, the expressions above may only be used to
indicate the right order, but not high accurateness, of the
expected injection.

Since the injected energy scales as $B^{-2}$ and the initial $ms$
periods are {\it required} for field amplification to operate, it
follows that the timescale for a substantial energy injection
(that is, an energy comparable to the kinetic energy of the
explosion itself, adopted to be $10^{51} erg$) is in fact very
short, of the order of $\sim hours$ (a ``normal'' pulsar would do
so in $\sim 100-1000 yr$). Therefore we may consider the injection
as quite instantaneous on astronomical explosion standards. In
short, the very formation of the magnetar leads to expect that a
large fraction of the kinetic energy of the remnant would be in
fact provided by the injection (see also Woosley 2010), and thus
to a modified dynamical behavior at later times.

This injected energy will make a very young remnant to expand more
than the corresponding case without energy injection, making it
look older in the free-expansion phase (lasting for just $\sim 100
yr$. The reason for that feature is the form of the injection
term, later the same energy would affect the Sedov-Taylor phase,
when the internal energy of the gas inside the cavity $U$ picks up
a term, becoming

\begin{equation}\label{energy}
U \, = \, E - {\frac{9}{32}}M{\dot R}^{2} -
\frac{L_{0}}{t^{-1}+{\tau_{0}}^{-1}}
\end{equation}
\\
where $R$ the radius of the SNR and $M$ the mass in motion, and
the last term represents the injected energy due to the magnetar
formation.

The results of these considerations were shown and discussed in
Allen \& Horvath (2004). A variety of works dealing with standard
SNR dynamics were then used to compare the results and point out
the differences (ex. Truelove \& McKee 1999, Luz \& Berry 1999,
van der Swaluw et al. 2001). Besides the mentioned quicker
expansion in the free-expansion phase, the Sedov-Taylor is also
modified to last longer than the case without injection, and
ending after $\sim 2 \times 10^{4} E_{51}^{3/14}n^{-4/7} \, yr$ as
a leading term (Allen \& Horvath 2004), when the SNR enters a
``snowplow'' (radiative) phase, not very relevant to our problem
because of the young ages expected. After recalling these main
results, we believe there are good reasons to keep in mind them
when SGR-AXP are tentatively associated with SNRs. A specific
search for these effects was undertaken by Vink \& Kuiper (2006),
with negative results. In our view, this is not surprising: after
$\sim \, 1000 \, yr$ the speed of the ejecta is essentially the
same for models with or without energy injection (Fig. 2 of Allen
\& Horvath 2004), making a kinematical identification more
difficult. Therefore it is only for very young remnants that a
sensible difference could be found. There are reports in the
literature (see Nomoto et al. 2010 and references therein) of
highly energetic supernovae (termed ``hypernovae'') featuring
$E_{ej} \geq 10^{52} erg$, but these have been identified with the
progenitors of GRBs, not necessarily the same events as the birth
of magnetars (however, see Yu, Cheng \& Cao 2010 for a unifying
model). Even in the magnetar-birth events, there are reasons to
believe that the total energy output is not large (Dall'Osso,
Shore \& Stella 2009), and therefore the failure to identify
hypernovae around SGR-AXP is, in principle, quite justified.

\section{The CTB 37B-CXOU J171405.7-381031 association}
\label{sect:CTB}

The suggestion of the association of CXOU J171405.7-381031 with
the remnant CTB 37B was made some time ago in Halpern \& Gotthelf
(2010a), and offers a new opportunity to understand the birth of
magnetars and their supernovae. These association has now been
confirmed by the measurements of the $\dot{P}$ that has been
measured by Sato et al. (2010) and Halpern \& Gotthelf (2010b),
qualifying the central object as a magnetar. Moreover, the
characteristic age is $\sim 1000 yr$ within a very small range, as
found by the two groups. A full discussion of the CTB 37B SNRs has
been addressed in several works (Aahronian et al. 2008a, 2008b,
Halpern \& Gotthelf 2010b). As in many other cases, the distance
to the remnant is uncertain ($10.2 \pm 3.5 kpc$; Caswell et al.
1975). It seems safe to assume, taking the angular diameter and
the smallest value of the distance range, that $R > 10 pc$, and
possibly a figure as big as $20 pc$ for the largest distance
scale.

Conventional models of SNR expansion (Truelove \& McKee 1999) run
into trouble to explain such radii for an age of only $1000 yr$.
Typically, a thousand-year old remnant will not enter the
Sedov-Taylor phase before $\sim 1400 yr$ unless the ejected energy
is larger than expected, but in any case this would have a small
effect on the radius since the latter scales as $R_{ST} = 15 pc
{(E_{ej}/{10^51} erg)^{1/5}} {(n_{ISM}/1 cm^{-3})^{-1/5}}
{(t/10^{4} yr)^{2/3}}$, with $n_{ISM}$ the particle density of the
interstellar medium in which the SNR expands and $t$ its age. It
follows that the remnant can be larger than $\sim 10 pc$, but only
well after $1000 yr$. This is at odds with the derived age of CXOU
J171405.7-381031 unless the actual radius happens to be $\sim 5
pc$ because a factor of $\geq 2$ error in the distance, which
seems unlikely. Fig. 1 displays the Radius-Age expected from SNRs
for the two cases, a ``standard'' expansion and a
``magnetar-driven'' expansion, for the same energy $10^{51} erg$
and assuming a value of $n_{ISM} = 1 cm^{-3}$ for the particle
density of the ISM.

As it can be checked from Fig. 1, the new data (Sato et al. 2010,
Halpern \& Gotthelf 2010b) leads to a young age for the
``magnetar-driven'' case only (upper dotted curve), and in fact
also favors the shorter of the estimated distances. Insistence on
the ``standard'' expansion scenario would need a much shorter
distance to the source, as stated above, and also the remnant to
be in the free-expansion phase. Since there is information
available about the density in the work of Aharonian et al.
(2008b), $n_{ISM} \sim 0.5 cm^{-3}$ introducing a negligible shift
in the curve (in fact, indistinguishable in a log plot), we
believe it is fair to consider the upper curve of Fig. 1 as
accurate and conclude that a magnetar-driven scenario for the
expansion is implied. In other words, a low density of the ISM can
not be invoked to explain the large radius of the remnant within a
``standard'' explosion, hence the solution of the quandary points
towards the dynamics univocally.

   \begin{figure}[h!!!]
   \centering
\includegraphics[height=.4\textheight]{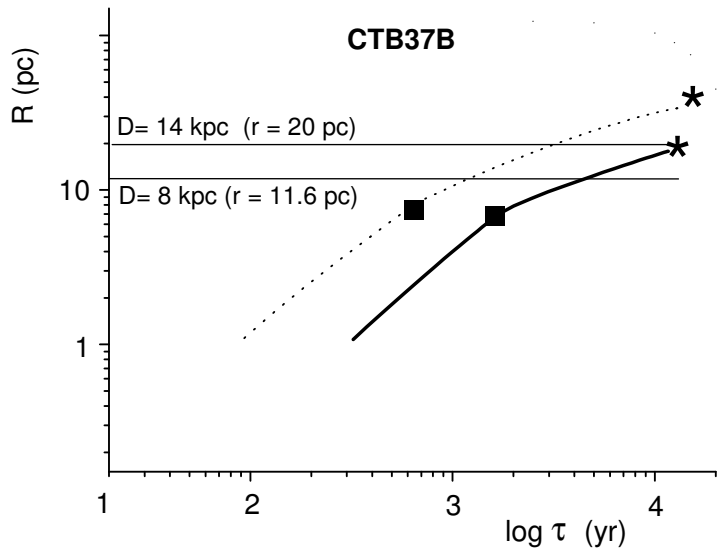}

   \caption{ Expansion of ``standard'' (lower solid curve) and ``magnetar-driven''
   (upper dotted curve) remnants.
   The two curves correspond to identical $E_{ej} = 10^{51} erg$ and external densities $n_{ISM} = 1 cm^{-3}$.
   As explained in the text and in Allen \& Horvath (2004), the difference in the solutions leads
   to entering before in the Sedov-Taylor phase in the latter case, and therefore the remnants expand
   faster than in the standard case. The transition of the free-expansion to the Sedov-Taylor phase is
   marked with a black square on each curve. The end of the
   Sedov-Taylor phase happens at the points marked with an
   asterisk.
   This is important to match the CTB 37B age with the derived $\tau = 1000 yr$
   of the associated magnetar CXOU J171405.7-381031, which acts as a relative chronometer. It can be
   checked that the matching is
   quite difficult for the ``standard'' expansion, or if the distance is closer to the extreme upper limit
   derived by Caswell et al. (1975) (horizontal lines, labelled).}

   \label{Fig1}
   \end{figure}

\section{Discussion and Conclusions}
\label{sect:conclusion}

We have discussed the association of the magnetar candidate CXOU
J171405.7-381031 with the CTB 37B remnant, with the aim of
contributing to the identification of progenitor of these exotic
objects, since the latter are still largely unknown. An initial
expectation of high-mass progenitors ($> 30-40 M_{\odot}$)
Gaensler et al. (2005), later supported by the cluster analysis by
Figer et al. (2005), Muno et al. (2005) and Bibby et al. 2008; has
recently been challenged by the identification of SGR 1900+14 by
Davies et al. (2009) with the Cl 1900+14 cluster, implying a
progenitor of $\leq 17 M_{\odot}$ for the cluster magnetar. If the
interpretation of these observations holds, the difference between
the events producing pulsars and magnetars should {\it not} be
related to the mass of the progenitor, but to some other
physical/evolutionary feature(s).

Our view of the problem is that, given the difficulties for
growing magnetic fields to the $10^{14}-10^{15} G$ scale, the
dynamo scenarios should operate. They lead in turn to predict a
large energy injection from the central object into the remnant,
and therefore a dynamical behavior of the SNR which has to be
considered with care. In other words, due to the injection of
energy by the central object (a combined effect of the high
rotation and the growing field, as required by $\alpha-\Omega$
dynamos) we do not deal with ordinary remnants, but rather with a
very special variety of them, the magnetar-driven ones. It is in
this framework that the association needs to be analyzed, as
reflected not only by the finding of an energetic magnetar inside,
but also by the difficulties of a standard SNR expansion of only
$\sim 1000 yr$ to match the observed features. A self-consistency
arises from the presence of a magnetar born at the explosion: its
presence reinforces a modified dynamics employment and explains
its size at a young age. Note that this is achieved with the {\it
same} total energy as a ``standard'' event, namely $10^{51} erg$.
If a larger scale is ever detected around a magnetar, its
expansion would dramatically show effects of the modified
dynamics, and would allow a refined test of this scenario.

Regarding the energy injection calculated to power the modified
dynamics, it could be much smaller if, for example, the rapidly
spinning object could get rid of its energy by gravitational
radiation or other form of ``invisible'' (i.e not coupled to the
remnant) emission. In spite that quadrupole GW would not compete
with dipole losses unless the oblateness happens to be extreme
(Allen \& Horvath 2004), and that it now agreed that $r$-mode
excitation is not important in these situations (Watts \&
Andersson 2002 , Rezzolla et al. 2001), this possibility can not
be completely discarded. However, if this possibility is realized,
we would learn that the birth of magnetars should be strong GW
burst sources and the problem of how to amplify the magnetic field
would be back.

Using the same reasoning, we have inferred before (Allen \&
Horvath 2011) a range of ages for CTB 37B much smaller than the
values derived within conventional models. Refining now the broad
interval to include a ISM density closer to usual values (as
directly measured for CTB 37B by Aharonian et al. 2008b), the
numbers for the ages are still low and would not change much
unless the mass of the progenitor was 3-4 times the $10 M_{\odot}$
value, but at the price of increasing the disagreement with the
characteristic age of CXOU J171405.7-381031. Note the age also
favors a ``light'' ($\sim 10 M_{\odot}$) progenitor. Moreover, the
new age is closer to the old estimation of $\sim 1500 yr$ by Clark
\& Stephenson (1975), but for quite different reasons. This new
age allowed a prediction of the $\dot P$ value, $\dot P \, \sim \,
4 \times 10^{-11} s \, s^{-1}$, which in turn predicted a magnetic
field strength of $B = 4 \times 10^{14} G$, {\it before} the
actual discovery to CXOU J171405.7-381031. It is also interesting
to note that younger objects ease the requirements for energizing
the TeV scale, as observed by the HESS Collaboration (Aharonian et
al. 2008b): electrons have to ``live'' less without being cooled
(Halpern \& Gotthelf 2010a) or even that the SNRs themselves
contribute to the emission, because they are actually younger than
they seem when their ages are erroneously estimated from
conventional SNR expansion models.

\normalem
\begin{acknowledgements}
We acknowledge the financial support received from the Funda\c
c\~ao de Amparo \`a Pesquisa do Estado de S\~ao Paulo. J.E.H.
wishes to acknowledge the CNPq Agency (Brazil) for partial
financial support.
\end{acknowledgements}

\label{lastpage}

\end{document}